\documentclass[aps,prl,twocolumn,noshowpacs,superscriptaddress]{revtex4-2}
\usepackage{graphicx,epsfig, subfigure}
\usepackage{amsbsy}
\usepackage{amssymb}
\usepackage{amsmath}
\usepackage{hyperref}
\usepackage[bottom]{footmisc}
\usepackage{xcolor}
\usepackage{color}
\usepackage{times}
\usepackage{textcomp}
\usepackage{verbatim} 

\date{\today}

\begin{document}

\newcommand{\eqnref}[1]{Eq.~\ref{#1}}
\newcommand{\figref}[2][]{Fig.~\ref{#2}#1}
\newcommand{\RN}[1]{%
  \textup{\uppercase\expandafter{\romannumeral#1}}%
}

\title{\Large  Switching and amplifying three-body Casimir effects}

\author{Zhujing Xu}
	\affiliation{Department of Physics and Astronomy, Purdue University, West Lafayette, Indiana 47907, USA}
\author{Peng Ju}
	\affiliation{Department of Physics and Astronomy, Purdue University, West Lafayette, Indiana 47907, USA}
\author{Xingyu Gao}
	\affiliation{Department of Physics and Astronomy, Purdue University, West Lafayette, Indiana 47907, USA}
	\author{Kunhong Shen}
	\affiliation{Department of Physics and Astronomy, Purdue University, West Lafayette, Indiana 47907, USA}
	\author{Zubin Jacob}
	\affiliation{Elmore Family School of Electrical and Computer Engineering, Purdue University, West Lafayette, Indiana 47907, USA}
	\affiliation{Birck Nanotechnology Center, Purdue University, West Lafayette, Indiana 47907, USA}	
\author{Tongcang Li}
	\email{tcli@purdue.edu}
	\affiliation{Department of Physics and Astronomy, Purdue University, West Lafayette, Indiana 47907, USA}
	\affiliation{Elmore Family School of Electrical and Computer Engineering, Purdue University, West Lafayette, Indiana 47907, USA}
	\affiliation{Birck Nanotechnology Center, Purdue University, West Lafayette, Indiana 47907, USA}	
	\affiliation{Purdue Quantum Science and Engineering Institute, Purdue University, West Lafayette, Indiana 47907, USA}
	\date{\today}

\begin{abstract}
 {\normalsize The dynamics of three interacting objects has been investigated extensively in Newtonian gravitational physics (often termed the three-body problem), and is important for many quantum systems, including nuclei, Efimov states, and frustrated spin systems. However, the dynamics of three macroscopic objects interacting through quantum vacuum fluctuations (virtual photons) is still an unexplored frontier. Here, we report the first observation of Casimir interactions between three isolated macroscopic objects. We propose and demonstrate a three terminal switchable architecture exploiting opto-mechanical Casimir interactions that can lay the foundations of a Casimir transistor.   Beyond the paradigm of Casimir forces between two objects in different geometries,  our Casimir transistor represents an important development for control of three-body virtual photon interactions  and will have potential applications in sensing and information
processing with the Casimir effect.  
 
 }
\end{abstract}

\maketitle

The interaction between three objects give rise to many fascinating phenomena such as chaos of astronomical objects \cite{Musielak_2014}, Efimov bound states of ultracold atoms \cite{kraemer2006evidence}, and frustrated states of quantum spin systems \cite{lacroix2011introduction}. It is intriguing to consider the potential of three-body interactions arising solely from quantum vacuum fluctuations (virtual photons) \cite{Casimir1948,RevModPhys.88.045003,GongCorradoMahbubSheldenMunday+2021+523+536}. 
The Casimir effect due to virtual photons can provide a new approach to couple  mechanical resonators \cite{PhysRevLett.122.030402}. Different from optomechanical coupling with real photons in cavity optomechanics \cite{xu2016topological,yang2020phonon,RevModPhys.86.1391,barzanjeh2021optomechanics}, optomechanical coupling with virtual photons will not suffer from cavity loss  and thus will not require a high-quality cavity. 
Recently, the Casimir effect was used to increase the quality factor of a mechanical resonator \cite{Pate2020CasimirSpring} and couple two separate mechanical resonators \cite{Fong2019,xu2021}. In addition, the Casimir effect  has  been utilized to realize nonlinear oscillation \cite{Chan2001nonlinear},  quantum trapping and self-assembling \cite{Zhao984,munkhbat2021tunable}.  
While the paradigm of Casimir effect between two objects has been extensively explored \cite{Sparnaay1958,PhysRevLett.88.041804,Lamoreaux1997,PhysRevLett.81.4549,Chan1941,Munday2009,PhysRevLett.120.040401,Tang2017}, the Casimir force between three macroscopic objects has not been detected yet. Beyond its fundamental interest, a Casimir system with three objects can open the route to realize crucial technological building blocks such as a transistor-like three-terminal device with quantum vacuum fluctuations. 


\begin{figure*}
	\centerline{\includegraphics[width=1.0\linewidth]{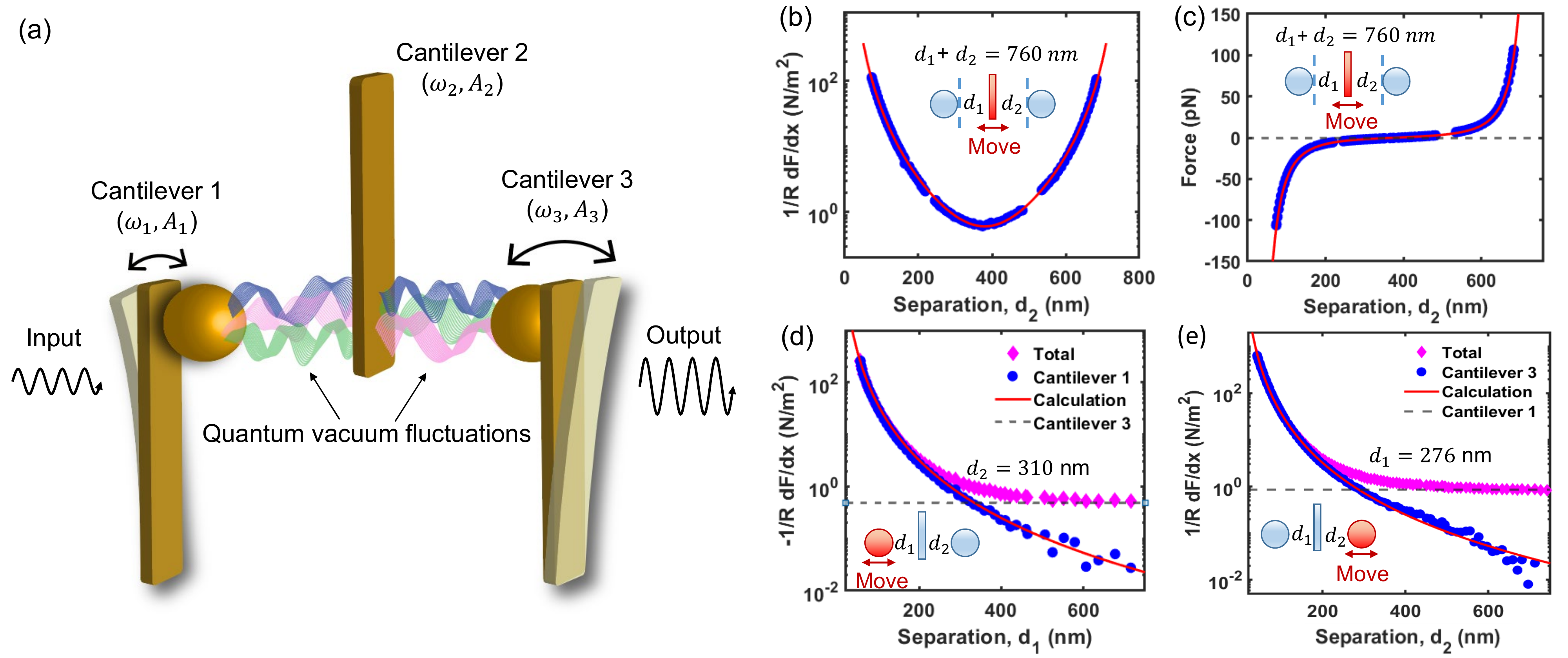}}
	\caption{\textbf{Casimir interaction between three optomechanical resonators.} (a) Three modified cantilevers with resonant frequencies $\omega_1$, $\omega_2$ and $\omega_3$ experience Casimir force between each two nearby surfaces. The vibration amplitudes of three cantilevers are denoted as $A_1$, $A_2$ and $A_3$. Additional parametric modulations are applied on the center cantilever to couple them by the Casimir effect. We can switch on and off the Casimir coupling between cantilever 1 and cantilever 3 by controlling the parametric modulations. In addition, we can  amplify the energy transfer through Casimir effect by adding an extra gain to cantilever 2.  (b). Measured Casimir force gradient on cantilever 2 (center) as a function of its position when the other two surfaces are fixed such that $d_1+d_2 = 760 $~nm. (c). The measured Casimir force on cantilever 2 is shown as a function of $d_2$. (d). Measured Casimir force gradient experienced by cantilever 2 as a function of $d_1$ when $d_2$ is fixed at $310$ nm. The red diamonds are the total force gradient $-\frac{1}{R}\frac{dF}{dx}$ measured from cantilever 2. The blue circles are the force gradient contributed from cantilever 1.  The red solid curve corresponds to the interaction between cantilever 1 and 2. The gray dashed line corresponds to the interaction between cantilever 2 and 3 and hence it is independent of $d_1$ under additivity approximation.  (e). Measured Casimir force gradient on cantilever 2 as a function of $d_2$ when $d_1$ is fixed at $276$ nm. }
	\label{Measurement}
\end{figure*}

In this article, we propose and demonstrate the first three-body Casimir system that can switch and amplify quantum-vacuum-mediated energy transfer, in analogy to a field effect transistor. 
Our unique three-body Casimir system consists of three closely-spaced optomechanical oscillators, as shown in Fig.\ref{Measurement}(a). Their motions are monitored by three independent fiber-optic interferometers. There are random quantum vacuum fluctuations between them and hence each cantilever experiences a separation-dependent Casimir force. We first measure the Casimir force between three objects. We then apply parametric modulation on cantilever 2 to couple their motion by the Casimir effect. In this way, energy can flow from cantilever 1 to cantilever 2 and  to cantilever 3. The center cantilever serves as a gate for controlling the energy transfer through the Casimir effect. By adding gain to the center cantilever with active feedback, we also realize amplification of the quantum-fluctuation-mediated energy transfer.  Our  Casimir transistor will have promising application in sensing \cite{javor2022zeptometer,Javor2021} and information processing \cite{PhysRevA.90.052313,Liu2016-ox}.

We first measure the Casimir force in our sphere-plate-sphere system (Fig.\ref{Measurement}). Assuming three surfaces are all made of ideal conductive metal and they are sufficiently thick, the Casimir force on the center one is \cite{Chan1941}
\begin{equation}
F^{0}_{2,C} = \frac{\pi^3\hbar c}{360}(\frac{R_1}{d_1^3}-\frac{R_2}{d_2^3}),
\label{Casimir_eq}
\end{equation} 
where $d_1$ and $d_2$ are the separation between cantilever 1 and cantilever 2, and the separation between cantilever 2 and cantilever 3 as shown in the inset of Fig.\ref{Measurement}.(b). $R_1$ and $R_2$ are the radii of the sphere on cantilever 1 and cantilever 3, respectively. 
The Casimir interaction between real materials can be calculated by the Lifshitz theory \cite{Lifshitz:1956,xu2021}.
We use the dynamic force measurement scheme to measure the Casimir force. More details about the calculation and measurement can be found in Methods and  Supplementary Information.

The measured Casimir force gradient on cantilever 2 in our three-body system is shown in Fig \ref{Measurement}.(b). We fix the position of cantilever 1 and 3 such that $d_1+d_2 = 760$ nm. Meanwhile, we change the position of the cantilever 2 (center). 
As the center cantilever moves from left side to the right side, the gradient meets the lowest value when $d_1 = d_2$ if $R_1=R_2$. At this specific separation, the net Casimir force on cantilever 2 is zero. 
The calculation based on Lifshitz's formula and proximity force approximation  is shown in the solid red curve. The measurement is in good agreement with the calculation. 
We also show the measured Casimir force gradient on cantilever 2 when separation $d_1$ is changed by moving cantilever 1 in Fig.\ref{Measurement}.(d), and similarly when 
separation $d_2$ is changed by moving cantilever 3 in Fig.\ref{Measurement}.(e).
While there have been many studies of Casimir interaction between two objects, our work reports the first  measurement of the Casimir force between three separate objects. It opens up the possibility for studying Casimir interaction between more complicated configurations, and can study the nonadditivity nature \cite{Messina2014,Milton2015} of the Casimir interaction by reducing the thickness of the center plate (see Supplementary Information for more details). 

\begin{figure*}
	\centerline{\includegraphics[width=0.90\linewidth]{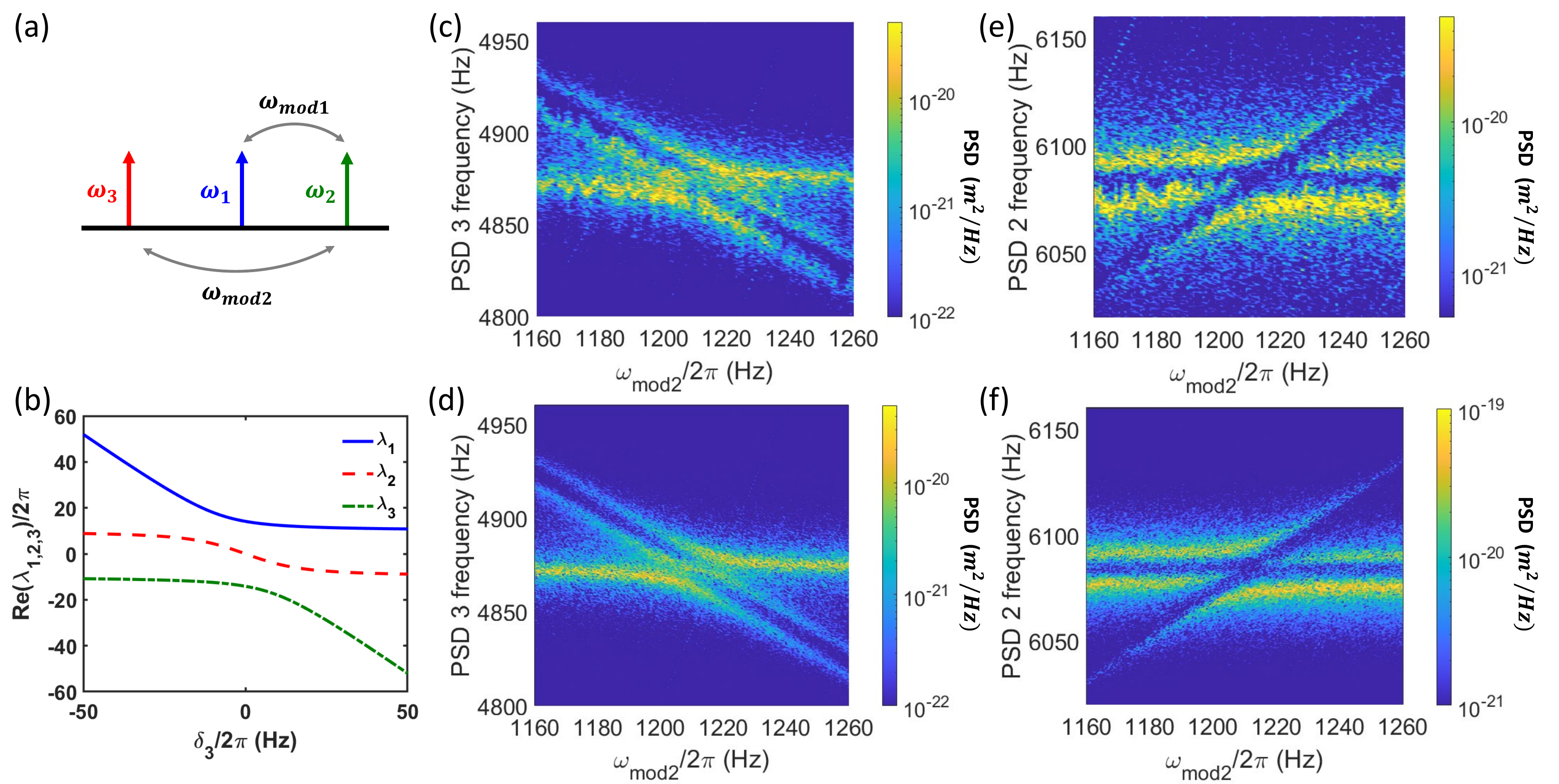}}
	\caption{\textbf{Coupling the vibrations of three cantilevers with the Casimir effect.} (a). Parametric modulation of the Casimir interaction is applied in our system. When $\omega_{mod1} = \omega_2-\omega_1$, cantilever 1 and cantilever 2 are coupled. Similarly, cantilever 2 and cantilever 3 are coupled when $\omega_{mod2} = \omega_2-\omega_3$. (b). Three eigenvalues of the Hamiltonian in Eq.(2) as a function of $\delta_3$ when $\delta_2 = 0$ and $|g_{12}| = |g_{23}| = 2\pi\times 20$~Hz. (c)  Power spectrum density (PSD) of cantilever 3 as a function of the modulation frequency $\omega_{mod2}$. (e) PSD of cantilever 2 as a function of $\omega_{mod2}$. The modulation amplitudes are $\delta_{d1} = 10.4$ nm and $\delta_{d2} = 14.1$ nm. The modulation frequency $\omega_{mod1}$ is fixed at 440 Hz. (d) and (f). The simulated PSD for two cantilevers. The separations are $d_1 = 88$ nm and $d_2 = 90$ nm.  }
	\label{Eigens}
\end{figure*}

We now use the Casimir effect to efficiently couple the motions of three cantilevers for  realizing  a more advanced Casimir-based device. The natural 
 frequencies and damping rates of three cantilevers are $\omega_1 = 2\pi\times 5661$ Hz, $\omega_2 = 2\pi\times 6172$ Hz,  $\omega_3 = 2\pi\times 4892$ Hz,  $\gamma_1 = 2\pi\times3.22$ Hz, $\gamma_2 = 2\pi\times 6.06$ Hz, and $\gamma_3 = 2\pi\times 3.58$ Hz when they are far apart. These frequencies shift under Casimir interaction.
The direct Casimir coupling strength between three cantilevers is smaller than the frequency differences between them. To solve this issue,
we use parametric coupling \cite{Huang2013,xu2021} by modulating the separation between each two cantilevers at a slow rate $\omega_{mod1,2}$ and a modulation amplitude $\delta_{d1,2}$. This is achieved by changing the position of the cantilever 2 as $\delta_{d1}\cos(\omega_{mod1} t)+\delta_{d2}\cos(\omega_{mod2} t)$. 
Such parametric modulation effectively couples three cantilever when $\omega_{mod1} = |\omega_1-\omega_2|$ and  $\omega_{mod2} = |\omega_3-\omega_2|$, as shown in Fig.\ref{Eigens}.(a).
Different from direct coupling that requires identical resonant frequencies, parametric coupling provides more freedom to couple different resonators. 
Under the parametric coupling scheme, the simplified Hamiltonian of the three-body system in the interaction picture  is (see Methods and
Supplementary Information for its derivation) \cite{xu2021}:
\begin{equation}
H = \begin{pmatrix} -i\frac{\gamma_1}{2} & \frac{g_{12}}{2} & 0\\ \frac{g_{12}}{2} & -i\frac{\gamma_2}{2}-\delta_2 & \frac{g_{23}}{2} \\ 0 & \frac{g_{23}}{2} & -i\frac{\gamma_3}{2}-\delta_3 \end{pmatrix}.
\label{eigenvalue}
\end{equation}%
where $\gamma_{1,2,3}$ denote the damping rates of the three cantilevers. $g_{12} = \frac{\Lambda_1}{2\sqrt{m_1m_2\omega_1\omega_2}}$ and $g_{23}=\frac{\Lambda_2}{2\sqrt{m_2m_3\omega_2\omega_3}}$ are the coupling strengths between cantilever 1 and cantilever 2, and between cantilever 2 and cantilever 3, respectively. Here we have $\Lambda_1 = \frac{d^2F_C(d_1)}{dx^2}\delta_{d1}$ and  $\Lambda_2 = \frac{d^2F_C(d_2)}{dx^2}\delta_{d2}$. $\delta_2 = \omega_1+\omega_{mod1}-\omega_2$ and $\delta_3 = \omega_1+\omega_{mod1}-\omega_{mod2}-\omega_3$ are the detuning of the system which depend on the modulation frequencies. The eigenvalues of this Hamiltonian near resonant coupling conditions are shown in Fig.\ref{Eigens}.(b). We can observe a clear two-fold anti-crossing when the detunings $\delta_3 = \delta_2= 0$.

\begin{figure*}
	\centerline{\includegraphics[width=0.9\linewidth]{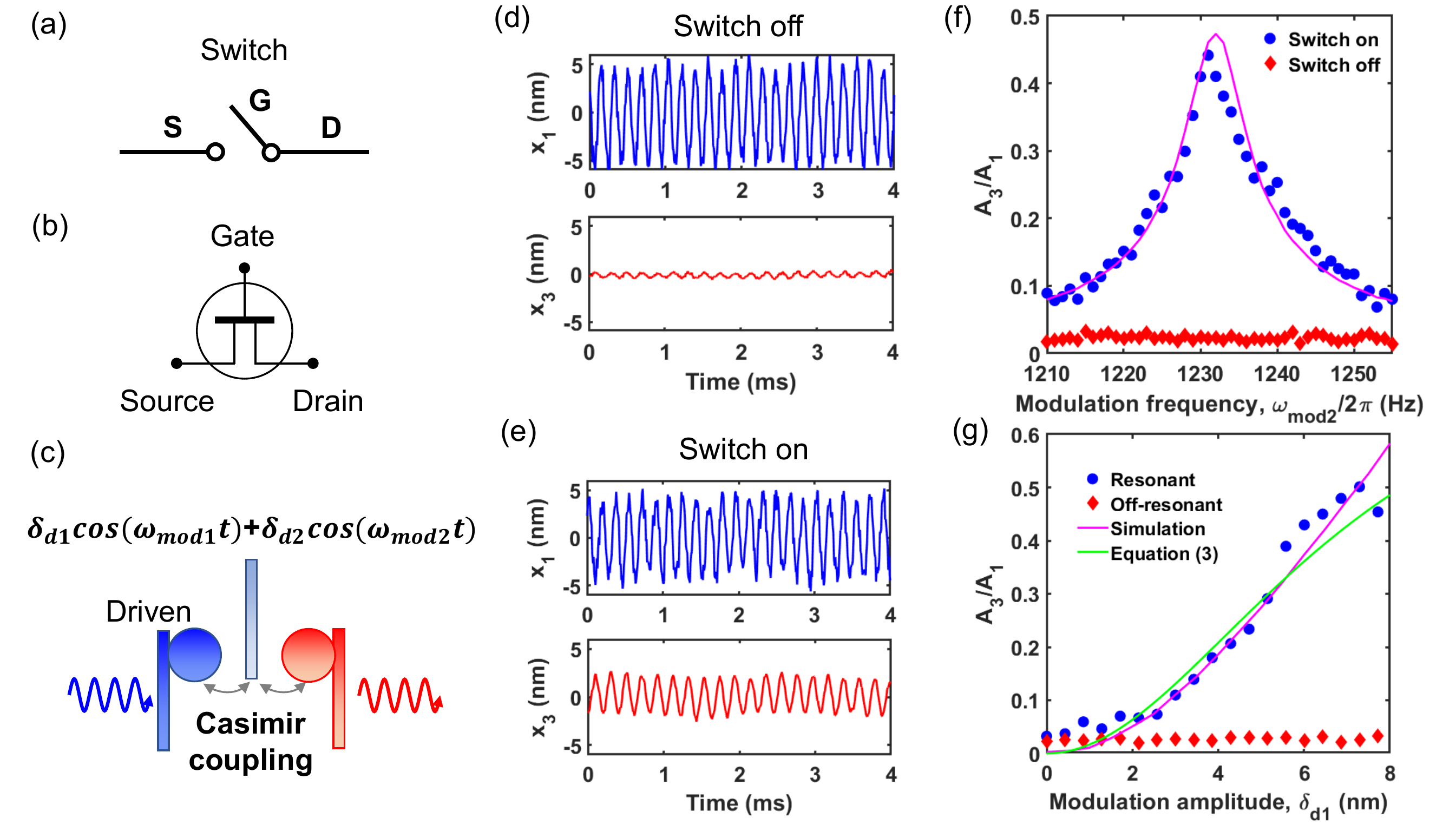}}
	\caption{\textbf{Switching quantum-fluctuation-mediated energy transfer.} (a). A symbolic switch. (b). A symbolic field effect transistor. (c). The quantum-fluctuation-mediated energy transfer between cantilever 1 and 3 can be switched on and off by  the modulation on cantilever 2.  (d). Measured displacement of two cantilevers when modulation is off. (e). Measured displacement of two cantilevers when modulation is on. Energy from cantilever 1 is transferred efficiently to cantilever 3. Here $\omega_{mod1} = 2\pi\times 465$ Hz,  $\omega_{mod2} = 2\pi\times 1230$ Hz, $\delta_{d1} = 6.0$ nm, and $\delta_{d2} = 8.5$ nm. The separations are $d_1 = 100$ nm and $d_2 = 105$ nm. (f). The transduction ratio is shown as a function of the modulation frequency $\omega_{mod2}$ when $\omega_{mod1}$ is on resonant.  (g). The transduction ratio as a function of modulation amplitude $\delta_{d1}$ when $\omega_{mod2} = 2\pi\times 1231$ Hz (on resonant, blue dots) and  $\omega_{mod2} = 2\pi\times 1150$ Hz (off resonant, red diamonds). $\omega_{mod1}$ is on resonant for both cases.
	}
	\label{Switch}
\end{figure*}

\begin{figure*}[t]
	\centerline{\includegraphics[width=0.9\linewidth]{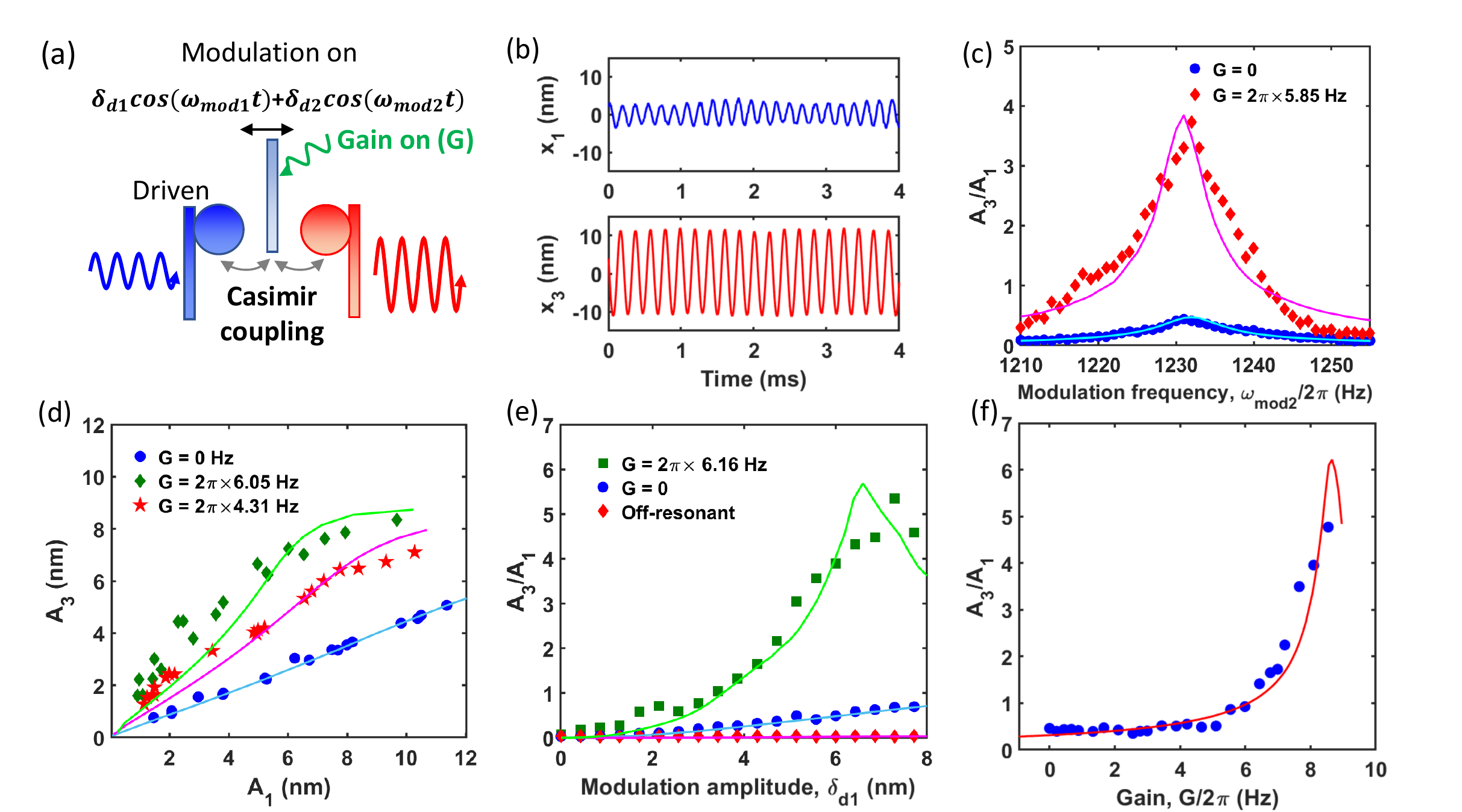}}
	\caption{\textbf{Amplifying quantum-fluctuation-mediated energy transfer.} (a). An extra gain is applied on cantilever 2 by feedback control together with parametric modulation. (b). The signal on cantilever 1 is transmitted to cantilever 3 with amplification. (c). The transduction ratio $A_3/A_1$ is shown as a function of modulation frequency $\omega_{mod2}$ when the gain is on and off.  $\omega_{mod1}$ is on resonant for both cases. (d). The amplitude $A_3$ is shown as a function of amplitude $A_1$ for three different gain coefficients. The parametric modulation is applied resonantly at the same time. (e). The ratio $A_3/A_1$ is shown as a function of modulation amplitude $\delta_{d1}$ for cases with gain, no gain, and off-resonant modulation. $\delta_{d2} = 1.42\delta_{d1}$. (f). The transduction ratio $A_3/A_1$ is shown as a function of the  extra feedback gain $G$ applied on cantilever 2.}
	\label{Amplify}
\end{figure*}

Our experimental results of the level repulsion behavior due to the Casimir coupling between three cantilevers are show in Fig.\ref{Eigens}.(c) and (e). We study this  behaviour experimentally by scanning the power spectrum densities (PSD) of cantilever 3 (Fig.\ref{Eigens}.(c)) and cantilever 2 (Fig.\ref{Eigens}.(e)) as a function of the  modulation frequency $\omega_{mod2}$ when $\omega_{mod1} = |\omega_1-\omega_2|$. Fig.\ref{Eigens}.(e) shows three  branches which correspond to the hybrid modes of three cantilevers after being projected to cantilever 2. Since $\omega_{mod1}$ is fixed at the resonant value that can couple cantilever 1 and cantilever 2, we notice a clear anti-crossing behavior (a horizontal dark line around 6080 Hz) independent of $\omega_{mod2}$. The two horizontal branches describe the coupled motion of cantilevers 1 and 2 . When we vary  $\omega_{mod2}$, we also observe an inclined branch with a frequency $\omega_3+\omega_{mod2}$ which corresponds to the motion of cantilever 3. When this inclined branch intersects with the other two branches at $\omega_{mod2} = |\omega_3-\omega_2|$, a more complicated level repulsion is observed. One mode disappears in the PSD of cantilever 2 as this mode only involves the motion of cantilever 1 and 3. 
More detailed discussion about the eigenvalues and PSD of the system is included in the Supplementary Information. 
Numerical simulation results are shown in Fig.\ref{Eigens}.(d) and (f), which agree well with experimental results.  Thus we have strongly coupled the motions of three objects with quantum vacuum fluctuations. 


Our three-body Casimir system enables switching (Fig.\ref{Switch}.(a)) and amplifying quantum-fluctuation-mediated energy transfer in analogy to a field effect transistor (Fig.\ref{Switch}.(b)). The quantum-fluctuation-mediated energy transfer between cantilever 1 and 3 can be easily switched  on and off by controlling the modulation on cantilever 2 (Fig.\ref{Switch}.(c)). 
When $\omega_{mod1}$ and $\omega_{mod2}$ are on resonance, vibration energy  from cantilever 1 can be transferred to cantilever 3 efficiently (Fig.\ref{Switch}.(e)). 
 However, when the modulation is off, the excitation on cantilever 1  can not be transferred to cantilever 3 efficiently (Fig. \ref{Switch}.(d)). 
In Fig.\ref{Switch}.(f), the measured amplitude ratio $A_3/A_1$ is shown as a function of modulation frequency $\omega_{mod2}$ for both switch on and off cases. We notice that the amplitude ratio can achieve up to 0.44 when the modulation is on resonance, and close to zero when the modulation if off resonance. Thus we can switch on and off the quantum-fluctuation-mediated energy transfer  with high contrast.

Under the steady state when the cantilever 1 is driven with a small amplitude and the parametric modulation on cantilever 2 is on resonant, the transduction ratio $A_3/A_1$ in this three-body Casimir system is (see Methods):
\begin{equation}
\frac{A_3}{A_1}= |\frac{\Lambda_1\Lambda_2}{4m_2m_3\omega_2\omega_3\gamma_2\gamma_3+\Lambda_2^2}|.
\label{ratio}
\end{equation}
where $\Lambda_1 = \frac{d^2F_C(d_1)}{dx^2}\delta_{d1}$ and  $\Lambda_2 = \frac{d^2F_C(d_2)}{dx^2}\delta_{d2}$. 
In Fig.\ref{Switch}.(g), the measured transduction ratio $A_3/A_1$ is shown as a function of modulation amplitude $\delta_{d1}$ when $\delta_{d2} = 1.42\delta_{d1}$. The transduction ratio is close to zero  for the off-resonant case. 
As expected, the ratio $A_3/A_1$ increases when $\delta_{d1}$ increases under resonant coupling. 
Our experimental results agree well with Eq.\ref{ratio} and numerical simulation results (Fig.\ref{Switch}.(g)).   


To realize a Casimir transistor with high efficiency, we introduce an extra gain to the system  (Fig.\ref{Amplify}.(a)) to amplify the quantum-fluctuation-mediated energy transfer.  
The extra gain is applied to cantilever 2 by feedback control such that the damping rate of cantilever 2 becomes $\gamma_2 = \gamma_{20}-G$, where $\gamma_{20}$ is the natural damping rate of cantilever 2 and $G$ is the gain coefficient (More details can be found in the Supplementary information). $\gamma_2$ becomes negative when $G>\gamma_{20}$. Based on Eq.\ref{ratio}, the transduction ratio $A_3/A_1$ increases when $\gamma_2$ decreases. 
Under such condition, energy from cantilever 1 is first transferred to cantilever 2 and get amplified and then transferred to cantilever 3. 
For example, we apply a fixed gain to cantilever 2 such that $G = 2\pi\times 8.73$ Hz to realize the amplification of energy transfer, as shown in Fig.\ref{Amplify}.(b). Other parameters are the same as those in Fig.\ref{Switch}.(e).

Figure \ref{Amplify}.(c) shows amplification of quantum-fluctuation-mediated energy transfer with our Casimir transistor.
 When a gain is applied to cantilever 2,
energy transfer from cantilever 1 to cantilever 3 shows a similar resonant behavior as the no-gain case, but has a striking improvement by a factor of 8 on the transduction ratio . The additional gain improves the quantum-fluctuation-mediated energy transfer efficiency  significantly. 
As expected, the transduction ratio $A_3/A_1$ increases when the parametric modulation amplitude (Fig.\ref{Amplify}.(e)) or the gain coefficient (Fig.\ref{Amplify}.(d),(f)) increases until the  system becomes unstable when the modulation amplitude or the gain is too large. 
Thus we have demonstrated amplification in a three-body Casimir system. The amplification function will be crucial for future applications of Casimir-based devices. For example, Casimir parametric amplification has been theoretically proposed for zeptometer  metrology \cite{javor2022zeptometer} and ultrasensitive magnetic gradiometry at the $10^{-18}$ T/cm level \cite{Javor2021}. 


In conclusion, we have measured the Casimir interaction between three objects, and demonstrated efficient coupling of three optomechanical resonators with virtual photons for the first time. Compared to the conventional optomechanical coupling with real photons in a high-Q cavity \cite{xu2016topological,yang2020phonon}, optomechanical coupling with virtual photons \cite{PhysRevLett.122.030402,PhysRevX.8.011031} does not need a high-Q cavity. 
Inspired by a field effect transistor, we also demonstrate switching and amplifying quantum-fluctuation-mediated energy transfer in our three-body Casimir system. As proposed by former theoretical studies, Casimir-based amplification and switching  will have applications in sensing \cite{javor2022zeptometer,Javor2021} and information processing \cite{PhysRevA.90.052313,Liu2016-ox}.


%

\newpage

\section{Methods}
\textbf{Casimir force calculation.}
At a finite temperature, the Casimir interaction comes from both quantum and thermal fluctuations.
At temperature T and separation x, the Casimir energy per unit area between two surfaces is given by \cite{Lifshitz:1956}
\begin{eqnarray}
E(x,T) = \frac{k_BT}{2\pi}\sum_{l=0}^{\infty}{}^{'}\int_{0}^{\infty} k_{\perp}dk_{\perp}\{\ln [1-r_{TM}^2(i\xi_l, k_{\perp})e^{-2xq}]\nonumber\\
+\ln[1-r_{TE}^2(i\xi_l,k_{\perp})e^{-2xq}]\}\hspace{1cm},
\label{CasimirEnergy}
\end{eqnarray}
where $\xi_l=\frac{2\pi k_B T l}{\hbar}$ is the Matsubara frequency and $k_{\perp} = \sqrt{k_x^2+k_y^2}$ is the wave vector parallel to the surface.
$r_{TE}(i\xi_l,k_{\perp})$ and $r_{TM}(i\xi_l,k_{\perp})$ are reflection coefficients of the transverse-electric and transverse-magnetic modes.
The separation between two surfaces is far smaller than the dimensions of the cantilever and the sphere. 
Therefore, we can apply the proximity-force approximation and the Casimir force between a sphere with radius $R$ and a plate is $F_C(x,T) = -2\pi RE(x,T)$. 
The calculation in \cite{xu2021} has shown that the contribution from thermal fluctuations at room temperature is less than 4$\%$ when the separation is less than 800 nm  Thus, the Casimir interaction in our system is dominated by quantum vacuum fluctuations.
In our system, the thickness of the center cantilever is 1 $\mu$m and the typical separation in our measurement is from 50 nm to 800 nm. 
Under such condition, the contribution from the nonadditivity is negligible compared to the sum of the pair potential and hence we take the additivity approximation \cite{Messina2014}. Under the thermal equilibrium, the force on the center cantilever can be simplified as $F_{2,C} = -F_{C} (d_1,T)+F_{C} (d_2,T)$. Under the additivity approximation, the force gradient between cantilever 1 and cantilever 2 is calculated by subtracting the force gradient between cantilever 2 and cantilever 3 from the total gradient experienced by cantilever 2, as shown in Fig.\ref{Measurement}.(d). 

\textbf{Experimental setup and force measurement.}
In the experiment, we use three modified AFM cantilevers to build the three-body Casimir system. The left and right cantilever has a dimension of $450\times 50\times 2$ $\mu$m$^3$. The center cantilever has a dimension of $500\times 100\times 1$ $\mu$m$^3$. Two 70-$\mu$m-diameter polystyrene spheres are attached to the free end of the left and right cantilevers to create the sphere-plate-sphere geometry. Additional 100-nm-thick gold layers are coated on both the sphere and cantilever surfaces.

During the measurement, we use phase-lock loop (PLL) to track the resonant frequency in the presence of the Casimir interaction. 
Then we can get the force gradient as $\frac{dF}{dx} = -2k\frac{\delta\omega}{\omega}$, where $k$ is the spring constant of the cantilever, $\delta\omega$ is the frequency shift in the presence of the interaction and $\omega$ is the natural resonant frequency. The separation between each two surfaces is calibrated by the electrostatic force. The frequency shift due to the electrostatic force and the Casimir force is $\Delta\omega = -\frac{\omega}{2k}\frac{\pi\epsilon_0R}{x^2}[(V_{ext}-V_c)^2+V_{rms}^2]-\frac{\omega}{2k}\frac{dF_C}{dx}$, where $V_{ext}$ is the external voltage applied on the surface, $V_c$ is the patch potential, $V_{rms}$ is the rms voltage fluctuations. $\frac{dF_C}{dx}$ is the force gradient of the Casimir interaction at separation $x$. By measuring the frequency shift of the cantilever for different external voltage $V_{ext}$, we can calculate the real separation between two surfaces. Our measurements show that the contribution from the rms voltage fluctuations is negligible compared to the Casimir force.
After canceling the contribution from electrostatic force, we can get the Casimir force gradient. The Casimir force gradient can be integrated over separation to obtain the Casimir force.

\textbf{Casimir force coupling and energy transfer.}
Under a slow modulation on cantilever 2, the separation between each two cantilevers is time-dependent such that 
\begin{eqnarray}
d_1(t) = d_{10}-\delta_{d1}\cos(\omega_{mod1}t)-\delta_{d2}\cos(\omega_{mod2}t)\nonumber\\
+x_1(t)-x_2(t),\nonumber\\
d_2(t) = d_{20}+\delta_{d1}\cos(\omega_{mod1}t)+\delta_{d2}\cos(\omega_{mod2}t)\nonumber\\
+x_2(t)-x_3(t).
\end{eqnarray}
Here $d_{10,20}$ is the equilibrium separation when there is no modulation applied, $\delta_{d1,d2}$ is the modulation amplitude, and $\omega_{mod1,2}$ are two modulation frequencies. $x_1(t)$, $x_2(t)$ and $x_3(t)$ describe vibrations of three cantilevers near their equilibrium positions. The motions of the cantilevers follow equations
\begin{eqnarray}
m_1\ddot{x_1}+m_1\gamma_1\dot{x_1}+m_1\omega_1^2x_1 = F_C(d_1(t))\hspace{2.3cm}\nonumber\\
m_2\ddot{x_2}+m_2\gamma_2\dot{x_2}+m_2\omega_2^2x_2 = -F_C(d_1(t))+F_C(d_2(t))\hspace{0.1cm}\nonumber\\
m_3\ddot{x_3}+m_3\gamma_3\dot{x_3}+m_3\omega_3^2x_3 = -F_C(d_2(t))\hspace{2cm}
\end{eqnarray}
Here we generalize the displacements $x_{1,2,3}(t)$ to complex values $z_{1,2,3}(t)$ such that $x_{1,2,3}(t) = Re[z_{1,2,3}(t)]$. We separate the fast-rotating term and the slow-varying term for $z_{1,2,3}(t)$ such that
\begin{equation}
z_{1,2,3}(t) = B_{1,2,3}(t)e^{-i\omega_{1,2,3}t},
\end{equation}
where $B_{1,2,3}(t)$ is the slow-varying amplitudes and we can neglect their second derivative terms $\ddot{B}_{1,2,3}(t)$ in the equations of motion. Under the limit of the small damping rate of three cantilevers such that $\gamma_{1,2,3}\ll\omega_{1,2,3}$ and the rotating wave approximation, 
the equation of motion can written as 
\begin{equation}
i\begin{pmatrix}\dot{B_1'}(t)\\ \dot{B_2'}(t)\\ \dot{B_3'}(t) \end{pmatrix} = \begin{pmatrix} -i\frac{\gamma_1}{2} & \frac{\Lambda_1}{4m_1\omega_1} & 0\\ \frac{\Lambda_1}{4m_2\omega_2} & -i\frac{\gamma_2}{2}-\delta_2 & \frac{\Lambda_2}{4m_2\omega_2} \\ 0 & \frac{\Lambda_2}{4m_3\omega_3} & -i\frac{\gamma_3}{2}-\delta_3 \end{pmatrix}\begin{pmatrix}B_1'(t) \\ B_2'(t)\\B_3'(t)\end{pmatrix},
\end{equation}
where $\Lambda_{1,2} = \frac{d^2F_C}{dx^2}|_{d_{01,02}}\delta_{d1,2}$. We have applied the transformation such that $B_1'(t) = B_1(t)$, $B_2'(t) = B_2(t)e^{i\delta_2 t}$, and $B_3'(t) = B_3(t)e^{i\delta_3 t}$, where $\delta_2 = \omega_1+\omega_{mod1}-\omega_2$ and $\delta_3 = \omega_1+\omega_{mod1}-\omega_{mod2}-\omega_3$ are the system detunings.
Under the steady condition, $\dot{B}_{1}$, $\dot{B}_{2}$, and $\dot{B}_{3}$ all equal to zero. The vibration amplitude of three cantilevers $A_{1,2,3}$ is the absolute value of the slow-varying component so we have $A_{1,2,3}(t) = |B_{1,2,3}(t)|$.
In this way, the ratio of $A_3/A_1$ is 
\begin{equation}
\frac{A_3}{A_1}=  | \frac{B_3}{B_1}| = |\frac{\Lambda_1\Lambda_2}{4m_2m_3\omega_2\omega_3\gamma_2\gamma_3+\Lambda_2^2}|.
\end{equation}
The vibrations of the three cantilevers can be quantized as phonons. By introducing normalized amplitudes $c_1 = \sqrt{\frac{m_1\omega_1}{\hbar}}B_1'$, $c_2 = \sqrt{\frac{m_2\omega_2}{\hbar}}B_2'$, and $c_3 = \sqrt{\frac{m_3\omega_3}{\hbar}}B_3'$, we obtain the equation of motion for the phonon modes as 
\begin{equation}
i\begin{pmatrix}\dot{c_1}\\\dot{c_2}\\\dot{c_3}\end{pmatrix} = \begin{pmatrix} -i\frac{\gamma_1}{2} & \frac{g_{12}}{2} & 0\\ \frac{g_{12}}{2} & -i\frac{\gamma_2}{2}-\delta_2 & \frac{g_{23}}{2} \\ 0 & \frac{g_{23}}{2} & -i\frac{\gamma_3}{2}-\delta_3 \end{pmatrix}\begin{pmatrix}c_1\\c_2\\c_3\end{pmatrix},
\end{equation}
where $g_{12} = \frac{\Lambda_1}{2\sqrt{m_1m_2\omega_1\omega_2}} = \frac{d^2F_C}{dx^2}|_{d_{01}}\delta_{d1}\frac{1}{2\sqrt{m_1m_2\omega_1\omega_2}}$, and $g_{23} = \frac{\Lambda_2}{2\sqrt{m_2m_3\omega_2\omega_3}} = \frac{d^2F_C}{dx^2}|_{d_{12}}\delta_{d2}\frac{1}{2\sqrt{m_2m_3\omega_2\omega_3}} $. 
Here we consider a special case that $g_{12} = g_{23}$, $\gamma_1 = \gamma_3$, and $\delta_{2,3} = 0$. The eigenvalues of the Hamiltonian are 
\begin{eqnarray}
\lambda_1 = -i\frac{\gamma_1}{2},\hspace{4.3cm}\nonumber\\
\lambda_2 = -i\frac{\gamma_1+\gamma_2}{4}+\frac{\sqrt{8g_{12}^2-(\gamma_1-\gamma_2)^2}}{4},\nonumber\\
\lambda_3 = -i\frac{\gamma_1+\gamma_2}{4}-\frac{\sqrt{8g_{12}^2-(\gamma_1-\gamma_2)^2}}{4}.
\end{eqnarray}

When the coupling strength is large compared to the damping difference such that $|g_{12}|> \frac{|\gamma_1-\gamma_2|}{2\sqrt{2}}$, we have $Im (\lambda_2) = -\frac{\gamma_1+\gamma_2}{4}$ and hence the steady state requires that 
\begin{equation}
\gamma_1+\gamma_2 >0.
\end{equation}
When the coupling strength is small compared to damping difference such that $|g_{12}|< \frac{|\gamma_1-\gamma_2|}{2\sqrt{2}}$, we have $Im (\lambda_2) = -\frac{\gamma_1+\gamma_2}{4}+\frac{\sqrt{(\gamma_1-\gamma_2)^2-8g_{12}^2}}{4}$. The steady state requires that 
\begin{equation}
\gamma_1+\gamma_2-\sqrt{(\gamma_1-\gamma_2)^2-8g_{12}^2}>0.
\end{equation} 





\end{document}